\documentclass[%
twocolumn,
letterpaper,
superscriptaddress,
amsmath,
amssymb,
prl,
reprint,
floatfix,
showkeys,
]{revtex4-2}
\setcitestyle{super}
\usepackage{graphicx}
\usepackage{siunitx}
\usepackage{textcomp,gensymb} 
\usepackage[hyperindex=true,colorlinks=true,linkcolor=blue,citecolor=blue,urlcolor=blue]{hyperref}
\usepackage{amsmath}
\usepackage{amssymb}

\draft 

\begin{document}
	
\title{Dynamics of nanosecond laser pulse propagation and of associated instabilities in a magnetized underdense plasma} 
	
\author{W. Yao}
\email{yao.weipeng@polytechnique.edu}
\affiliation{LULI - CNRS, CEA, UPMC Univ Paris 06 : Sorbonne Universit\'e, Ecole Polytechnique, Institut Polytechnique de Paris - F-91128 Palaiseau cedex, France}
\affiliation{Sorbonne Universit\'e, Observatoire de Paris, Universit\'e PSL, CNRS, LERMA, F-75005, Paris, France}

\author{A. Higginson}
\affiliation{Center for Energy Research, University of California San Diego, 9500 Gilman Drive, La Jolla, California 92093-0417, USA}

\author{J.-R Marqu\`{e}s}
\affiliation{LULI - CNRS, CEA, UPMC Univ Paris 06 : Sorbonne Universit\'e, Ecole Polytechnique, Institut Polytechnique de Paris - F-91128 Palaiseau cedex, France}

\author{P. Antici}
\affiliation{INRS-EMT, 1650 boul, Lionel-Boulet, Varennes, QC, J3X 1S2, Canada}

\author{J. B\'{e}ard}
\affiliation{CNRS, LNCMI, Univ Toulouse 3, INSA Toulouse, Univ Grenoble Alpes, EMFL, 31400 Toulouse, France}

\author{K. Burdonov}
\affiliation{LULI - CNRS, CEA, UPMC Univ Paris 06 : Sorbonne Universit\'e, Ecole Polytechnique, Institut Polytechnique de Paris - F-91128 Palaiseau cedex, France}
\affiliation{Sorbonne Universit\'e, Observatoire de Paris, Universit\'e PSL, CNRS, LERMA, F-75005, Paris, France}
\affiliation{JIHT, Russian Academy of Sciences, 125412, Moscow, Russia}	

\author{M. Borghesi}
\affiliation{School of Mathematics and Physics, The Queen’s University Belfast, Belfast, UK}

\author{A. Castan}
\affiliation{LULI - CNRS, CEA, UPMC Univ Paris 06 : Sorbonne Universit\'e, Ecole Polytechnique, Institut Polytechnique de Paris - F-91128 Palaiseau cedex, France}
\affiliation{CEA, DAM, DIF, F-91297}

\author{A. Ciardi}
\affiliation{Sorbonne Universit\'e, Observatoire de Paris, Universit\'e PSL, CNRS, LERMA, F-75005, Paris, France}

\author{B. Coleman}
\affiliation{School of Mathematics and Physics, The Queen’s University Belfast, Belfast, UK}

\author{S. N. Chen}
\affiliation{``Horia Hulubei'' National Institute for Physics and Nuclear Engineering, RO-077125 Bucharest-Magurele, Romania}

\author{E. d’Humières}
\affiliation{University of Bordeaux, CELIA, CNRS, CEA, UMR 5107, F-33405 Talence, France}

\author{T. Gangolf}
\affiliation{LULI - CNRS, CEA, UPMC Univ Paris 06 : Sorbonne Universit\'e, Ecole Polytechnique, Institut Polytechnique de Paris - F-91128 Palaiseau cedex, France}

\author{L. Gremillet}
\affiliation{CEA, DAM, DIF, F-91297}
\affiliation{Université Paris-Saclay, CEA, LMCE, 91680 Bruyères-le-Châtel, France}

\author{B. Khiar}
\affiliation{Office National d'Etudes et de Recherches Aérospatiales (ONERA), Palaiseau 91123, France}

\author{L. Lancia}
\affiliation{LULI - CNRS, CEA, UPMC Univ Paris 06 : Sorbonne Universit\'e, Ecole Polytechnique, Institut Polytechnique de Paris - F-91128 Palaiseau cedex, France}

\author{P. Loiseau}
\affiliation{CEA, DAM, DIF, F-91297}
\affiliation{Université Paris-Saclay, CEA, LMCE, 91680 Bruyères-le-Châtel, France}

\author{X. Ribeyre}
\affiliation{University of Bordeaux, CELIA, CNRS, CEA, UMR 5107, F-33405 Talence, France}

\author{A. Soloviev}
\affiliation{IAP-RAS, Nizhny Novgorod, Russia}	

 \author{M. Starodubtsev}
 \affiliation{IAP-RAS, Nizhny Novgorod, Russia}

\author{Q. Wang}
\affiliation{Institute of Applied Physics and Computational Mathematics, Beijing 100094, China}
\affiliation{Department of Electrical and Computer Engineering, University of Alberta, 9211 116 St. NW, Edmonton, Alberta T6G 1H9, Canada}

\author{J. Fuchs}
\email{julien.fuchs@polytechnique.edu}
\affiliation{LULI - CNRS, CEA, UPMC Univ Paris 06 : Sorbonne Universit\'e, Ecole Polytechnique, Institut Polytechnique de Paris - F-91128 Palaiseau cedex, France}

\date{\today}
	
\begin{abstract}

The propagation and energy coupling of intense laser beams in plasmas are critical issues in laser-driven inertial confinement fusion. Applying magnetic fields to such a setup has been evoked to enhance fuel confinement and heating, and mitigate laser energy losses. Here we report on experimental measurements demonstrating improved transmission and increased smoothing of a high-power laser beam propagating in an underdense magnetized plasma. We also measure enhanced backscattering, which our simulations show is due to hot electrons confinement, thus leading to reduced target preheating.

\end{abstract}

\pacs{}

\maketitle 


The propagation and energy coupling of intense laser pulses in underdense plasmas (defined as having electron density $n_e<n_c \equiv 10^{21}\lambda_{\mu m}^{-2}$, in units of $cm^{-3}$, where $\lambda_{\mu m}$ is the laser wavelength in $\mu$m) have been extensively researched, because of their paramount importance to laser-driven inertial confinement fusion (ICF) \cite{nuckolls1972laser,kruer1991intense,zylstra2022burning}. For ICF, it is critical that the maximum laser energy possible is coupled either directly to the fuel in direct-drive \cite{craxton2015direct} or to the hohlraum wall in indirect-drive \cite{lindl1995development} in a spatially homogeneous manner, as the laser's imprint seeds hydrodynamic instabilities that limit fuel compression \cite{casner2021recent}.
Laser-plasma interaction (LPI) can be either beneficial to ICF, e.g. when spatially smoothing the laser energy distribution \cite{fuchs2001experimental,malka2003enhanced}, or detrimental, e.g. by conversely causing spikes in the laser pattern through self-focusing \cite{pesme2002laser,lancia2011anomalous}, or by inducing energy loss through stimulated Raman and Brillouin scattering (SRS and SBS, respectively) \cite{montgomery2016two}. The former scattering mechanism can further induce, through the generation of forward hot electrons, detrimental preheating of the fuel, limiting the laser intensity used in ICF \cite{kruer1988physics}. 

In the pursuit of better performance in ICF experiments, applying external magnetic fields to ICF targets has been proposed in both direct and indirect-drive schemes \cite{chang2011fusion,perkins2017potential}, particularly in improving fuel temperature \cite{montgomery2015use} and reducing hydrodynamic instabilities \cite{perkins2017potential}. Additionally, magnetization also impacts laser propagation \cite{watkins2018magnetised} and LPI processes \cite{gong2015mitigating}.
Besides, in the scheme of magnetized liner inertial fusion (MagLIF), there is an increasing need for a detailed understanding of the axial magnetic field effects on the laser-plasma coupling via LPI processes  \cite{gomez2014experimental,shi2019three}.
For example, prior works have already investigated how a magnetic field parallel to the laser propagation affects the laser propagation and instabilities, both theoretically \cite{liu2018faraday,los2021magnetized} and experimentally \cite{montgomery2015use}. However, when the magnetic field is not simply parallel to the laser, there is no clear understanding \cite{hassoon2009stimulated,paknezhad2011nonlinear,winjum2018mitigation}, nor detailed experimental investigation, of its effects. 

In this paper, we experimentally explore the dynamics of a single speckle \cite{montgomery2000flow,wattellier2003generation,masson2014stimulated} laser beam propagating through pre-ionized underdense plasma targets. For this setup, a large-scale and strong (20 T) transverse magnetic field can be applied. This is used as a proxy for the indirect-drive ICF hohlraums environment. Primarily, we report on enhanced energy transmission and improved beam smoothing in a magnetized plasma, via time-resolved and two-dimensional (2D) space-resolved transverse imaging of the transmitted beam. Previously, measurements of the electron temperature and density indicated that the magnetized plasma is hotter \cite{froula2007quenching}, a result of reduced thermal transport across the magnetic field. This behavior is here reproduced by large-scale three-dimensional (3D) magneto-hydrodynamic (MHD) simulations. Additionally, measurements of the backscattered SRS indicate, at a low energy level of $10^{-5}$ of the incident laser, an increased signal level at different plasma densities in the magnetized case. 2D particle-in-cell (PIC) simulations confirm this trend and reveal that hot electron  confinement (with energies up to 100 keV) is responsible for this increase, which will reduce preheating of an ICF target.

\begin{figure}[hbtp]
\centering
\includegraphics[width=0.45\textwidth]{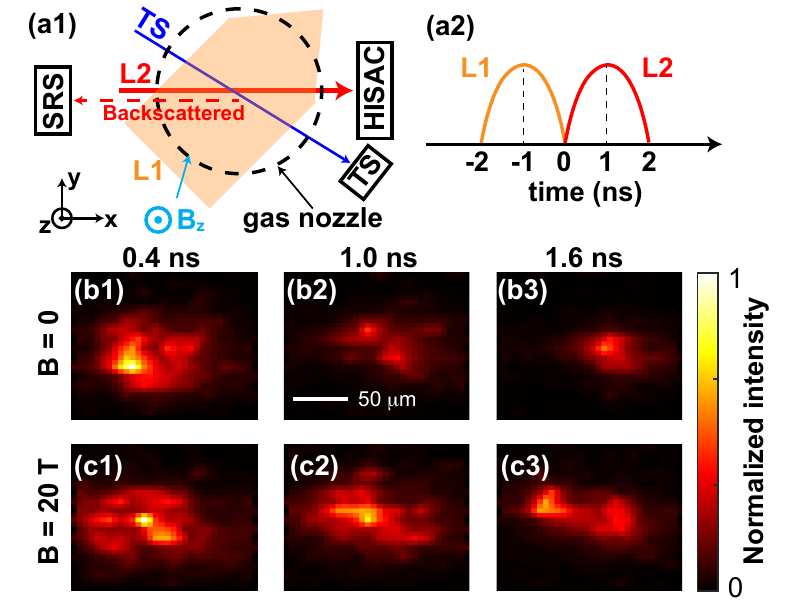}
\caption{(a1) Schematic displaying the experimental configuration (top view, see text for details). (a2) Time sequence of the pre-heating (L1) and interaction (L2) beams. (b-c) Temporal snapshots from the HISAC diagnostic, displaying transmitted L2 laser light, for a peak electron density $n_e = 0.04 n_c$ and magnetic field $B = 0$ and $20$ T, respectively. 
All panels are normalized by their respective maximum intensity and share the same colormap on the right.}
\label{exp_setup}
\end{figure} 

The experiment was performed at the Laboratoire pour l'Utilisation des Lasers Intenses LULI2000 facility, using the configuration shown in Fig.~\ref{exp_setup} (a). Two nanosecond-duration laser beams, both operating at a wavelength $\lambda_0 = 1.053$ $\mu$m and having Gaussian pulse duration $\tau_0 = 1$ ns full-width-at-half-maximum (FWHM), were used. A first laser beam (L1) was used to pre-ionize a hydrogen gas target, generated by a supersonic gas jet nozzle with an aperture diameter equal to 2 mm. L1 was focused to a large, FWHM spot size equal to 2$\times$0.3 mm$^2$ (horizontal and vertical, respectively), with an energy $E_0 = 30$ J; resulting in an intensity on target $I_0 = 3\times10^{12}$ W$\cdot$cm$^{-2}$. During the falling edge of L1, see Fig.~\ref{exp_setup} (a2), the main interaction beam (L2) is focused at the center of the fully ionized hydrogen plasma. L2 was polarized along the z-direction. It was focused using an f/22 lens into a single speckle of spot size $70\times70$ $\mu$m$^2$ (FWHM) and the Rayleigh length $Z_r \sim 2$ mm. It contained an energy $E_0 = 50$ J, resulting in an intensity at focus of $I_0 = 1.4\times10^{15}$ W$\cdot$cm$^{-2}$. 
Both lasers propagated at 0.75 mm above the nozzle opening.
The plasma electron density was varied by adjusting the backing pressure of the gas jet system. 
The electron number density at the center of the gas jet, which has a density profile of 1.5 mm (FWHM), was varied in the range $n_e = 0.02 - 0.08$ $n_c$.
The plasma was magnetized by a magnetic field directed along the positive z-axis, the same as the injected gas flow. This field was generated by a pulsed-power driven Helmholtz coil \cite{albertazzi2013production,higginson2017detailed}, with field strength $B = 20$ T. The field is steady-state ($> 100 \  \mu$s) and homogeneous ($\sim$ 1 cm) relative to the plasma dynamics and scale.

The L2 transmission was characterized by collecting the on-axis transmitted light using an f/10 lens. Its aperture is larger than that of the focusing lens (f/22). The focal spot of the laser was imaged onto a high-speed 2D spatially-resolved sampling camera (HISAC) composed of a fiber optics bundle coupled to a streak camera with a 30 ps temporal resolution \cite{kodama1999development,nakatsutsumi2010high}. 
Additionally, a probe beam with $\lambda_0 = 0.527$ $\mu$m, duration $\tau_0 = 1$ ns (FWHM), and focal spot 300 $\mu$m served to drive Thomson scattering (TS) of the plasma electron waves \cite{froula2007quenching}, allowing us to determine the electron number density and temperature, at the center of the focal spot of L2. 
The TS diagnostic was time-resolved via a streak camera coupled to a spectrometer, providing 30 ps resolution.  
Finally, the backscattered laser light due to SBS and SRS, collected within the full aperture of the L2 focusing optics, was recorded, using a time-resolved (via streak camera) and a time-integrated apparatus, respectively.

\begin{figure}[hbtp]
\centering
\includegraphics[width=0.45\textwidth]{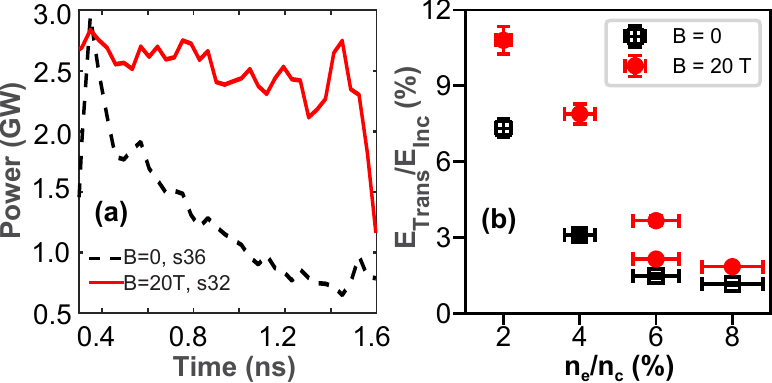}
\caption{(a) Power of the signal received by the HISAC diagnostic as a function of time, for a peak density of $n_e = 0.04 n_c$. 
(b) Time-integrated transmitted energy (normalized to the incident laser energy), measured with HISAC as a function of the background plasma peak electron density. Note that the density corresponds to the peak density of the fully ionised jet, based on calibration conducted off-line, with neutral gas. The horizontal error bars represent the calibration uncertainty, while the vertical error bars represent the noise level of the corresponding shots. 
}
\label{exp_HISAC}
\end{figure} 

We first discuss the increased laser transmission and smoothing achieved in the magnetized case. 
Figure~\ref{exp_setup} (b) and (c), as well as Fig.~\ref{exp_HISAC} summarize the HISAC measurements. An example of reconstructed temporal snapshots of the HISAC is shown in Fig.~\ref{exp_setup} (b) and (c), from 0.4 ns to 1.6 ns after the start of L2, for a peak density of $n_e = 0.04n_c$. For the unmagnetized case, the transmitted light signal recorded is clearly decreasing, both in strength (see Fig.~\ref{exp_HISAC}a) and size (see Fig.~\ref{exp_setup}b) with time. 
One of the possible reasons lies in the laser self-focusing, which will induce a considerable angular divergence, past the focusing point, as well as filamentation \cite{kruer1988physics,pesme2002laser,lancia2011anomalous} for the propagating light. Both can reduce the laser energy being able to be collected by the collecting optics. Strong self-focusing is expected in these experimental conditions. Its growth time, estimated by $\tau_{f} = 2\pi\gamma^{-1}\tau_0$, where $\gamma = 0.125 \left(v_{os}/v_{th}\right)^2 \omega_{pe}^2/\omega_0$ is the growth rate \cite{montgomery2016two}, $\tau_0$ and $\omega_0$ are the laser period and angular frequency, $v_{os}$ and $v_{th}$ are the electron oscillation and thermal velocities, respectively, is indeed 0.4 ps for $T_e = 100$  eV (as measured by TS), i.e., much smaller than our laser duration.  
However, for the magnetized case, as previous measurements\cite{froula2007quenching} and our simulations show, the electron density is lower and the plasma temperature is higher. With the growth rate of self-focusing being proportional to $\omega_{pe}^2 / v_{th}^2 \sim n_e / T_e$, it hints that the self-focusing will be weaker in the magnetized case, compared to the unmagnetized one.

This coincides with what we observe, for the magnetized case, i.e. that the transmitted light signal keeps both its strength (see Fig.~\ref{exp_HISAC}a) and finite width (see Fig.~\ref{exp_setup}c). 
Fig.~\ref{exp_HISAC} (a) shows the transmitted light power as a function of time for the peak density of $n_e = 0.04 n_c$. More energy is transmitted in the magnetized case, particularly at later times.
The fraction of transmitted laser light to incident light is shown in Fig.~\ref{exp_HISAC} (b) as a function of the peak density.
For the entire density range investigated, transmission is larger through the magnetized plasma than the unmagnetized one.

\begin{figure}[htp]
\centering
\includegraphics[width=0.45\textwidth]{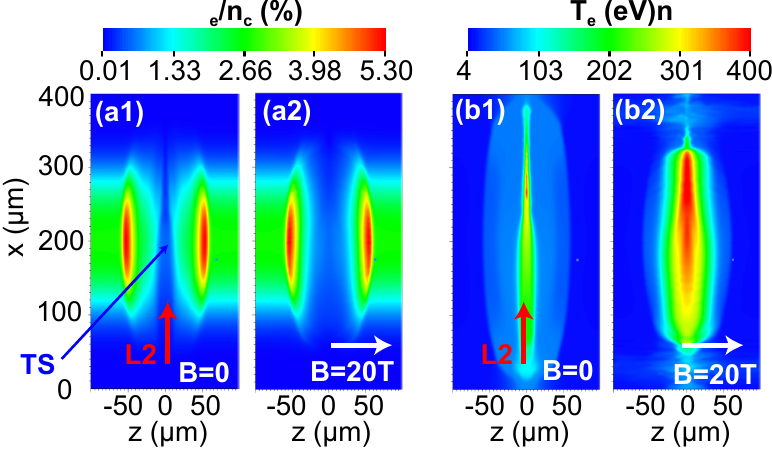}
\caption{
2D maps (of the 3D MHD simulation results using the experimental parameters) sliced in the xz-plane at $y = 0$ at $t = 1$ ns, showing the electron number density in (a1) and (a2), and the electron temperature in (b1) and (b2) for both $B =$ 0 and 20 T cases, respectively.
All results above are for the peak density of $n_e = 0.04 \ n_c$. }
\label{exp_TS}
\end{figure} 

The increased laser transmission in the magnetized case is favored not only by reduced self-focusing, as mentioned above, but also by a reduced absorption in the lowered density and hotter plasma, as we expect the plasma to be lower density and hotter in the magnetized case compared to the unmagnetized one \cite{froula2007quenching}.  

These two trends are consistent with the fact that thermal transport across a magnetic field can be substantially reduced, especially for electrons when their Hall parameter $H_e = \omega_{ce}\tau_{ei} > 1$, in which $\omega_{ce}$ is the electron cyclotron frequency and $\tau_{ei}$ is the electron-ion collision time. 
In our case, an estimate with $n_e \sim 0.02 \ n_c$ and $T_e \sim 100$ eV, as supported by our TS measurements and also shown in Fig.~\ref{exp_TS} (a) and (b), gives $H_e \approx 10$. This indicates strong thermal transport confinement perpendicular to the magnetic field. Note that since the thermal plasma beta $\beta = 8\pi n_e k_BT_e/B^2 \approx 5$, the external magnetic field does not affect the overall plasma dynamics except for the above transport process.
As a result of the plasma conditions in the magnetized case, the L2 laser experiences a lower inverse Bremsstrahlung absorption rate \cite{richardson20192019}, which is another possible explanation for the increased laser transmission observed in Fig.~\ref{exp_HISAC} (b).

To go beyond the above detailed analytical estimates, we perform 3D MHD simulations using the code FLASH \cite{fryxell2000flash} with the same parameters as in the experiment. 
For the magnetized case, anisotropic magnetized thermal diffusion (both for electrons and ions) is included. Details about the simulation setup can be found in the supplementary material.

The results of the simulated electron density distribution are shown in Fig.~\ref{exp_TS} (a1) and (a2). A sharper and narrower expansion gradient (along the z-direction) of the plasma and higher density within the laser path (located around $z=0$) in the unmagnetized case is apparent. The corresponding electron temperatures are shown in Fig.~\ref{exp_TS} (b1) and (b2). The plasma becomes hotter within the laser path and the expansion along the x-direction (i.e., perpendicular to the applied magnetic field) is more confined in the magnetized case, which is in agreement with the above experimental measurements. 

\begin{figure}[htp]
\centering
\includegraphics[width=0.45\textwidth]{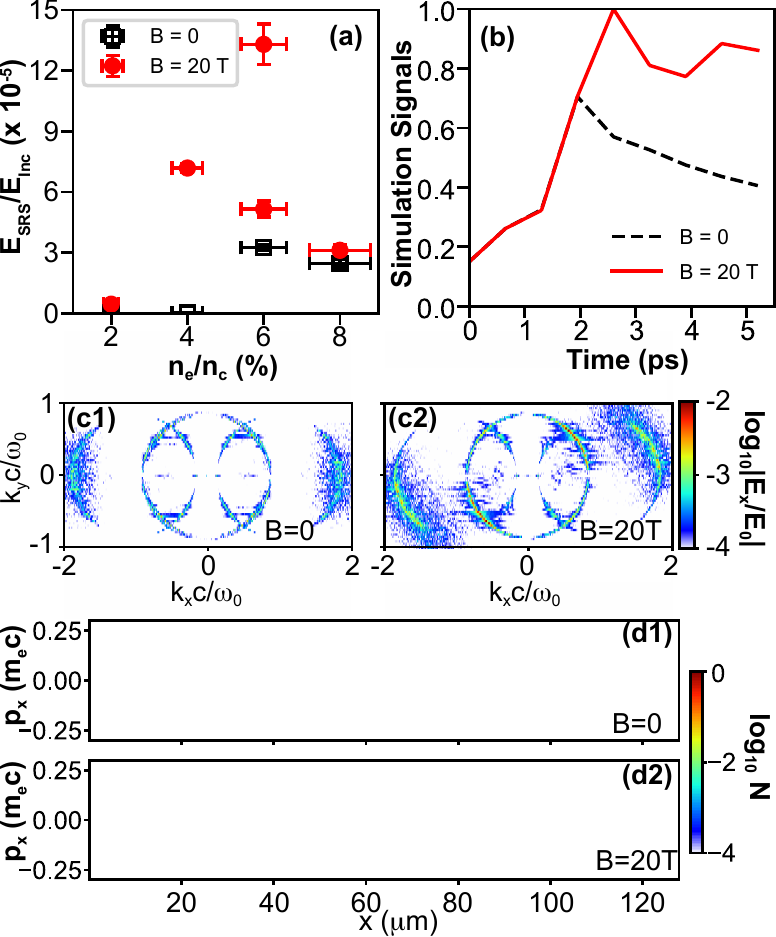}
\caption{(a) Experimentally measured backscattered SRS light energy (normalized by the incident laser energy) for various peak electron densities for cases with (red dots) and w/o (black triangle) the magnetic field, respectively. The horizontal error bars represent the calibration uncertainty, while the vertical error bars represent the noise level of the corresponding shots.
(b) 2D PIC simulation results for the time evolution of the SRS signals, using the initial parameters of $n_e = 0.02$ $n_c$, $T_e = 200$ eV, and a plane wave laser with $a_0 = 0.033$, normalized by the maximum value. The red solid line is for the magnetized case, while the black dashed line is for the unmagnetized one. (c) 2D FFT of the $E_x$ field and (d) the phase-space of $x$-$p_x$ for electrons are shown for cases with and w/o magnetic field, respectively. The color bar represents the normalized $E_x$ strength (to the incident electric field $E_0$) and particle number N in the logarithmic scale, respectively.
All the results are at $t = 5.2$ ps.}
\label{exp_SRS}
\end{figure}

We now discuss the results of the backscattered diagnostics. We first note that we did not observe significant changes in backscattered SBS between the magnetized and unmagnetized cases. However, this is not the case for SRS.
Figure~\ref{exp_SRS} (a) shows a summary of the backscattered SRS results under various density conditions for both the unmagnetized and magnetized cases. Note that the level of SRS is weak compared to the incident laser (mainly because of the low plasma density) at around $10^{-5}$. Yet, contrary to former theoretical calculations \cite{winjum2018mitigation,gong2015mitigating}, it is clear that the SRS signals here in our experiment are stronger in the presence of the 20 T external magnetic field. 

This is investigated by performing 2D PIC simulations with the code SMILEI \cite{derouillat2018smilei}.
Figure~\ref{exp_SRS} (b) shows that the results of the 2D (conducted in the xy-plane) PIC simulations (with $n_e = 0.02$ $n_c$, $T_e = 200$ eV, and a plane wave laser with $a_0 = eE/m_ec\omega_0 = 0.033$, where $E$, $\omega_0$, $e$, $m_e$, and $c$ are the laser electric field and frequency, electron charge and mass, and light speed, i.e., as in the experiment) are consistent with the experimental measurements.  Note that here the incident laser propagates along the x-direction and is $E_y$-polarized, while the applied magnetic field is along the z-direction. The magnetic field axis is consistent with that of the experiment (and thus the electron gyromotion is within the xy-plane of the simulation), but the laser is here polarized along the y-direction, in order for the electrons dynamics to be within the simulation plane. More details about the simulation setup can be found in the supplementary material.

Note that our parameters are different from those used in \citet{winjum2018mitigation} with $n_e \sim 0.13$ $n_c$ and $k\lambda_D \geq 0.3$ (SRS is in the strongly damped regime); 
in our case, owing to the low electron number density ($n_e = 0.02$ $n_c$), the backscattered SRS is quite weak, as mentioned above and as can be seen in the experimental data in Fig.~\ref{exp_SRS} (a). 
Moreover, from the three-wave coupling conditions and linear theory, we find that with our parameters, the backscattered SRS (with $k\lambda_D \sim 0.25$, within the kinetic regime \cite{kline2006different}) has a wavenumber of $k_x = 1.8$ in the 2D FFT of $E_x$ fields, as shown in Fig.~\ref{exp_SRS} (c1) and (c2); and that the growth rate is not affected by the magnetic field (as expected with $a_0 \ll 1$ and $\omega_{ce} \ll \omega_{pe}$), as shown in Fig.S3 (c).

We find that the reason for the enhanced SRS is rather the confinement of the hot electrons (trapped inside the Langmuir wave induced by the incident laser) due to the external magnetic field, which can be seen clearly from the $x$-$p_x$ phase-space distribution shown in Fig.~\ref{exp_SRS} (d1) and (d2), i.e., the backward drift of the electrons due to the $v\times B$ force with negative $p_x$ values. For the trapped superthermal electrons, with a velocity of 0.2c (c is the velocity of the light), their Larmor radius is around 20 $\mu$m and the cyclotron period is around 2 ps, which means that they can be well confined by the magnetic field within our system. This confinement behavior accounts for the sustained SRS signals in the magnetized case shown in Fig.~\ref{exp_SRS} (b); otherwise, the SRS signal will decrease as the unmagnetized one. Note that this could be potentially beneficial to ICF since the hot electrons that could preheat the target (with energy around 50 keV \cite{solodov2020hot}, i.e., corresponding to a Larmor radius around 30 $\mu$m) would be well confined by the magnetic fields.
Besides, this could also be beneficial to other exciting branches of research, such as SRS-based laser compression\cite{malkin1999fast} and amplification\cite{qu2017plasma,shi2018laser} schemes.

In summary, we report on experimental and simulation results systematically exploring laser beam propagation and energy coupling in a low-density magnetized plasma for the first time. With dedicated diagnostics, we find that the laser transmission is enhanced in the magnetized case, which is favored by a lower electron density and higher electron temperature, thus leading to a lower absorption (via inverse Bremsstrahlung). Moreover, we find that the transmitted light is less self-focused and more uniform, which is favorable in terms of illumination inhomogeneity of the final target.  3D MHD simulations qualitatively support the above measurements. We also find enhanced backscattered SRS in the magnetized case, which is well-reproduced by 2D PIC simulations and can be explained by the confinement of the hot electrons, which is another beneficial aspect brought by the external magnetic field to the ICF scheme through a lower preheating of the target. 


This work was supported by the European Research Council (ERC) under the European Union's Horizon 2020 research and innovation program (Grant Agreement No. 787539). The authors acknowledge the expertise of the LULI laser facility staff. The computational resources of this work were supported by the National Sciences and Engineering Research Council of Canada (NSERC) and Compute Canada (Job: pve-323-ac, PA). 


\bibliography{main}
\bibliographystyle{apsrev4-2-titles}
		
\end{document}


\title{Supplemental Material: Dynamics of nanosecond laser pulse propagation and of associated instabilities in magnetized underdense plasma}

\maketitle 




To investigate the underlying physics responsible for the clear parametric dependencies observed in the experiment, we incorporate a two-stages simulation process using two kinds of simulation codes. The codes in the first stage used in this process are the MHD codes FLASH \cite{fryxell2000flash} and GORGON \cite{ciardi2007evolution} which are used to calculate the ambient plasma electron temperature ($T_e$) and density ($n_e$) based on the laser and target parameters used experimentally and which are compared to the HISAC and TS measurements. The second stage is to use PIC simulations to elucidate the dynamics of SRS, using the code SMILEI \cite{derouillat2018smilei}.

\section{3D FLASH simulations}

\begin{figure*}[htp]
\centering
\includegraphics[width=\textwidth]{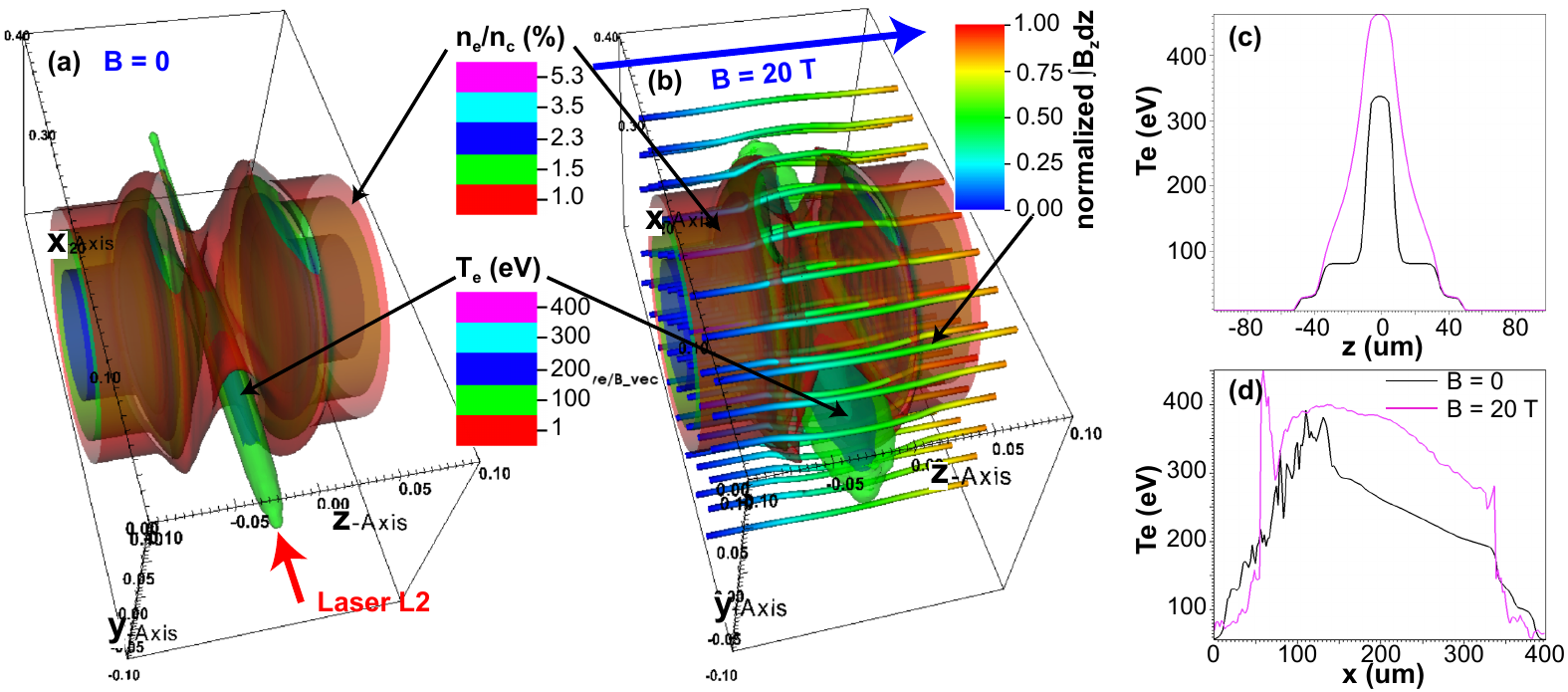}
\caption{3D MHD simulation results using the experimental parameters (including the temporal profile of L1 and L2) for a peak background electron density $n_e = 0.04 n_c$. (a) and (b) 3D rendering of the ambient electron number density, electron temperature, and magnetic field lines for both the unmagnetized and magnetized cases, respectively. It also demonstrates the simulation configuration, i.e., the gas jet and B-field are along the z-direction, while the laser L2 is along the x-direction. The lineouts of the electron temperature are detailed in (c) along the z-direction at $x = 200 \ mu$m and $y = 0$; and in (d) along the x-direction at $y = 0$ and $z = 0$, for both the $B =$ 0 and 20 T cases, respectively. 
All results are taken at $t = 1$ ns. 
}
\label{simu_FLASH}
\end{figure*} 

The results of the MHD FLASH simulations are shown in Fig.~\ref{simu_FLASH}. From the 3D rendering in Fig.~\ref{simu_FLASH} (a) and (b), we can see that the injected gas is in the z-direction, along which the magnetic field of 20 T is applied. It has an initial ambient electron number density $n_e = 0.04 n_c$ and the spatial profile is taken from the experimental calibration of the nozzle at 0.75 mm away from the opening, as in the experiment. Two lasers are used to model L1 and L2. We use in the simulation the same laser parameters and temporal profile as in the experiment. L2 is along the x-direction (illustrated by the red arrow). For clarity, L1 is not shown here.  
Magnetic field thermal diffusion (anisotropic for both ions and electrons) is included with a flux-limiter $f = 0.06$ \cite{gray1977observation,chatzopoulos2019gray}, as well as the Biermann battery effects; however, the Nernst term is ignored. For the radiation transport, multi-group diffusion with 29 groups is used. Laser energy deposition was done with ray-tracing and refraction by inverse Bremsstrahlung absorption.

Qualitatively, the simulation results show that with the external magnetic field along the z-direction, the electron thermal conductivity along the x-direction is reduced and the electron temperature becomes higher due to this confinement, as can be seen in Fig.~\ref{simu_FLASH} (d). Besides, the magnetic pressure is not negligible and enlarges the channel along the z-direction, as can be seen in Fig.~\ref{simu_FLASH} (c).

\section{2D PIC simulations}

\begin{figure}[htp]
\centering
\includegraphics{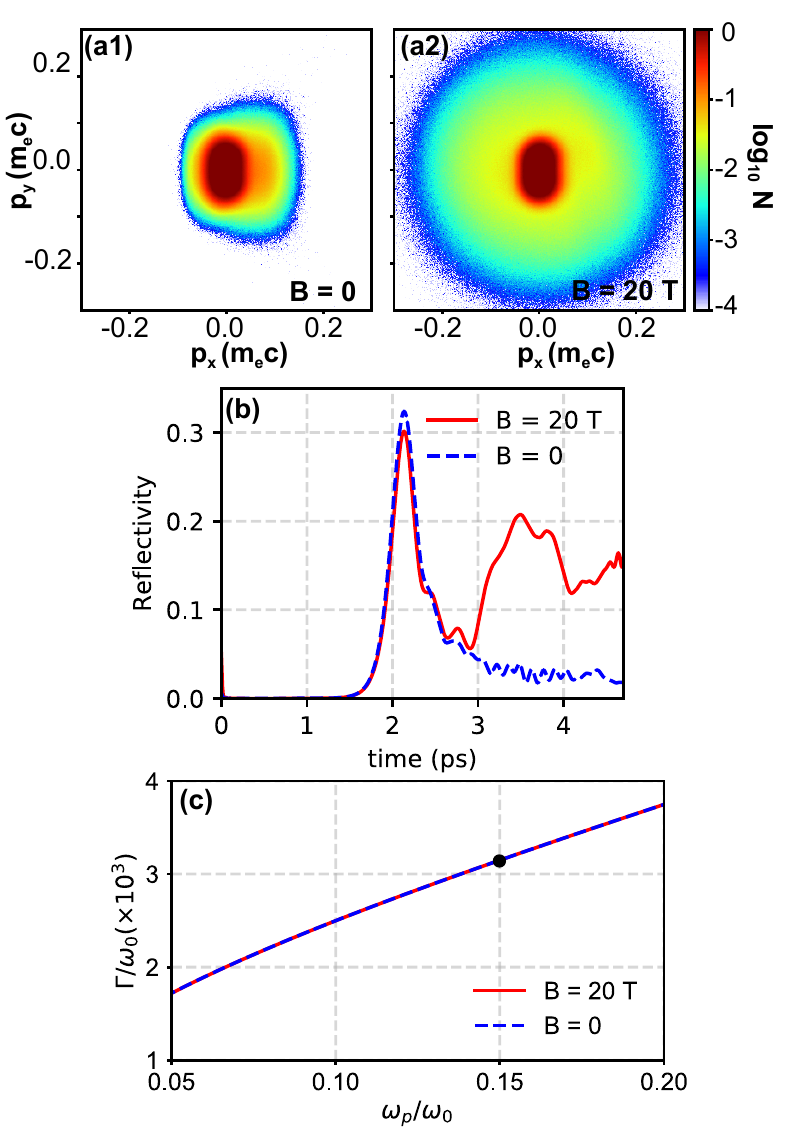}
\caption{2D PIC simulation results with $n_e = 0.02$ $n_c$, $T_e = 200$ eV and $a_0 = 0.03$. For the unmagnetized and magnetized case, respectively, the phase-space of $p_x$-$p_y$ (normalized by $m_ec$) are shown in (a1) and (a2). The color bar represents the normalized particle number N in the logarithmic scale.  All the results are at $t = 5.6$ ps. The temporal evolution of the SRS reflectivity for both cases is shown in (b). The SRS growth rate as a function of electron number density is shown in (c). The black dot indicates our parameter, i.e., $n_e = 0.02$ $n_c$. }
\label{simu_SMILEI}
\end{figure}

2D PIC simulations are used to shed light on the magnetized LPI kinetic process, using the code SMILEI \cite{derouillat2018smilei}.
Detailed results are shown in Fig.~\ref{simu_SMILEI}. In this simulation, the initial electrons have a temperature $T_e = 200$ eV and a number density $n_e = 0.02 n_c$. For backscattered SRS, it gives $k\lambda_D \approx 0.26$. 
The laser parameters we use are the ones of the experimental interaction beam L2, i.e., having a normalized field $a_0 = 0.03$. We use a plane wave along the y-direction because our simulation box size along y (i.e., $L_y = 32\ \mu$m) is smaller than the laser focal spot radius (i.e., 70 $\mu$m). The laser propagates along the x-direction.
The external magnetic field of 20 T is applied along the z-direction and the electron cyclotron motion is within the xy-plane, with a frequency of $\omega_{ce} = 0.013 \omega_{pe}$. With our initial conditions, the Larmor radius of an electron is $r_e \approx 2 \ \mu$m, while that of a proton is $r_i \approx 70 \ \mu$m. 
The cyclotron period for the electrons is $\tau_{ce} \approx 2$ ps. Note that the characteristic growth time for backscattered SRS is $\tau_{SRS} \approx 3.8$ ps, which is on the same order of $\tau_{ce}$. 
We use 100 particles per cell, with a grid of 4096 $\times$ 1024 cells, and a simulation box size of 128 $\times$ 32 $\mu$m$^2$. The simulation duration is around 6 ps. Boundary conditions along the x-direction are open for fields and thermal for particles, while along the y-direction they are periodic for both fields and particles.

Note that our parameters are different from those used in \citet{winjum2018mitigation}, for which SRS is in the strongly damped regime; whereas here, owing to the low electron number density ($n_e = 0.02$ $n_c$), the backscattered SRS is quite weak, as can be seen in the experimental data in Fig. 4 (a), i.e., $E_{SRS}/E_{inc} \sim 10^{-5}$ (where $E_{SRS}$ is the energy of the SRS signals while $E_{inc}$ is that of the incident laser). 

The underlying mechanism, behind the enhancement of the backscattered SRS signals with an external magnetic field, as is measured in our experiment, lies in the confinement of the hot electrons (generated inside the Langmuir wave induced by the incident laser) due to the external magnetic field. This is clearly seen from the $x$-$p_x$ phase-space distribution shown in Fig. 4 (d1) and (d2), as well as in the $p_x$-$p_y$ phase-space distribution shown in Fig.~\ref{simu_SMILEI} (a1) and (a2). The latter illustrates the backward drift of the electrons due to the $v\times B$ force with negative $p_x$ values. For the superthermal electrons generated from SRS, with a velocity above 0.2c (c is the velocity of the light), their Larmor radius is around 20 $\mu$m and the cyclotron period is around 2 ps, which means that they can be well confined by the magnetic field within our system. Indeed, in Fig.~\ref{simu_SMILEI} (a1) and (a2), we see that the hot electrons generated from SRS in the unmagnetized case propagate freely along the x-direction until they are out of the system (with lower momenta in both $p_x$ and $p_y$ within 0.2c). However, those in the magnetized case are confined inside the system due to cyclotron motion, with higher momenta reaching 0.25c in both directions. The latter behavior accounts for the sustained SRS signals in the magnetized case shown in Fig. 4 (b); otherwise, the SRS signal will decrease with time as the unmagnetized one. 

Moreover, the temporal evolution of the SRS reflectivity in Fig.~\ref{simu_SMILEI} (b) shows the same behavior, i.e., before the hot electrons leave the simulation box in the unmagnetized case around $L_x / p_x/m_e \sim 2.7$ ps, both cases have almost the same reflectivity. After that, the SRS reflectivity in the unmagnetized case drops quickly; while with the magnetic field confinement effect, it remains at a much higher level in the magnetized case, compared to the unmagnetized one. 

\begin{figure}[htp]
\centering
\includegraphics[width=0.4\textwidth]{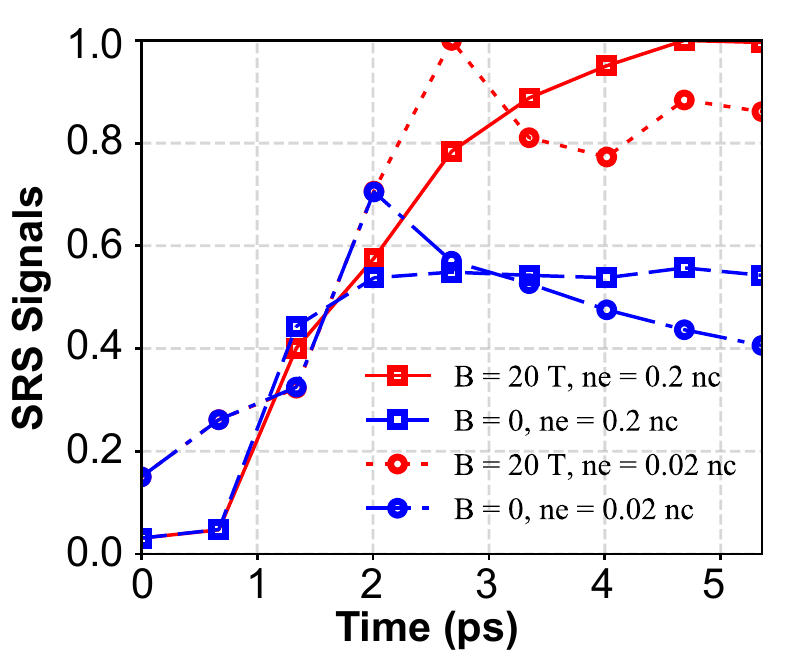}
\caption{Parametric study of backscattered SRS using 2D PIC simulations. The initial number density and temperature is $n_e = 0.02$ $n_c$ and $T_e = 200$ eV (dashed lines), as well as $n_e = 0.2$ $n_c$ and $T_e = 1$ keV (solid lines), for both the unmagnetized (blue) and magnetized (red) cases. All other initial conditions are the same.}
\label{simu_scan}
\end{figure}

These simulated SRS reflectivity results are consistent with the experimental measurement shown in Fig. 4 (a), as well as the integration of the SRS signals in the 2D FFT of $E_x$ shown in Fig. 4 (b). 
The similarity before 3 ps in both cases can be justified by the three-waves coupling conditions and linear theory. With our parameters, denoted by the black dot in Fig.~\ref{simu_SMILEI} (c), the magnetic field strength is not strong enough to affect the SRS growth rate (with $a_0 \ll 1$ and $\omega_{ce} \ll \omega_{pe}$). 
We have also checked that when the initial number density is 0.2 $n_c$ and the temperature is 1 keV (as shown in Fig.~\ref{simu_scan}), the essential B-field effect is the same. 

In addition, for the protons, their cyclotron period is above 3 ns. This is consistent with our benchmark simulations using mobile ions, from which it is clear that having mobile or immobile ions does not affect the conclusions. 
Besides, the ``ring''-like structure in the 2D FFT map of $E_x$ fields, see Fig. 4 (c), is related to the incident laser itself and the periodic boundary condition along the y-direction, more investigations about this will be done in the future.

\bibliography{refs}
\bibliographystyle{apsrev4-2-titles}